\documentclass[sigconf,screen,10pt,dvipsnames]{acmart}
\usepackage{xspace}
\usepackage{glossaries}
\glsdisablehyper

\newacronym{acpi}{ACPI}{Advanced Configuration and Power Interface}
\newacronym{afu}{AFU}{Accelerator Function Unit}
\newacronym{asic}{ASIC}{Application-Specific Integrated Circuit}
\newacronym{atf}{ATF}{ARM Trusted Firmware}
\newacronym{bdk}{BDK}{Board Development Kit}
\newacronym{bist}{BIST}{built-in self-test}
\newacronym{bmc}{BMC}{Baseboard Management Controller}
\newacronym{bram}{BRAM}{Block RAM}
\newacronym{capi}{CAPI}{Coherent Accelerator Processor Interface}
\newacronym{capp}{CAPP}{Coherent Accelerator Processor Proxy}
\newacronym{ccip}{CCI-P}{Core Cache Interface}
\newacronym{ccix}{CCIX}{Cache Coherent Interconnect for Accelerators}
\newacronym{cheri}{CHERI}{Capability Hardware Enhanced RISC Instructions}
\newacronym{cpld}{CPLD}{Complex Programmable Logic Device}
\newacronym{cxl}{CXL}{Compute eXpress Link}
\newacronym{dac}{DAC}{Digital Analog Converter}
\newacronym{dma}{DMA}{Direct Memory Access}
\newacronym{dpu}{DPU}{Data Processing Unit}
\newacronym{dsl}{DSL}{Domain-Specific Language}
\newacronym{eci}{ECI}{Enzian Coherence Interface}
\newacronym{ept}{EPT}{Extended Page Table}
\newacronym{fiu}{FUI}{FPGA Interface Unit}
\newacronym{fmc}{FMC}{FPGA Mezzanine Card}
\newacronym{fpga}{FPGA}{Field Programmable Gate Array}
\newacronym{gpu}{GPU}{Graphics Processing Unit}
\newacronym{hbm}{HBM}{High-Bandwidth Memory}
\newacronym{hpc}{HPC}{High-Performance Computing}
\newacronym{i2c}{I\textsuperscript{2}C}{Inter-Integrated Circuit}
\newacronym{ic}{IC}{Integrated Circuit}
\newacronym{iommu}{IOMMU}{I/O Memory Management Unit}
\newacronym{ipi}{IPI}{Inter-Processor Interrupt}
\newacronym{ipmi}{IPMI}{Intelligent Platform Management Interface}
\newacronym{ipu}{IPU}{Infrastructure Processing Unit}
\newacronym{isa}{ISA}{Instruction Set Architecture}
\newacronym{mmio}{MMIO}{Memory-Mapped I/O}
\newacronym{mmu}{MMU}{memory management unit}
\newacronym{mpsoc}{MPSoC}{Multiprocessor System-on-a-Chip}
\newacronym{mpx}{MPX}{Memory Protection Extensions}
\newacronym{msi}{MSI}{Message-Signaled Interrupt}
\newacronym{msr}{MSR}{Model-Specific Register}
\newacronym{nic}{NIC}{network interface adapter}
\newacronym{nvme}{NVMe}{NVM Express}
\newacronym{ocapi}{OpenCAPI}{Open Coherent Accelerator Processor Interface}
\newacronym{pae}{PAE}{Physical Address Extensions}
\newacronym{pcb}{PCB}{Printed Circuit Board}
\newacronym{pcie}{PCIe}{PCI Express}
\newacronym{pmbus}{PMBus}{Power Management Bus}
\newacronym{psci}{PSCI}{Power State Coordination Interface}
\newacronym{psl}{PSL}{POWER Service Layer}
\newacronym[longplural=page table entries]{pte}{PTE}{page table entry}
\newacronym{qpi}{QPI}{QuickPath Interconnect}
\newacronym{rdma}{RDMA}{Remote Direct Memory Access}
\newacronym{rot}{RoT}{Root-of-Trust}
\newacronym{rpc}{RPC}{Remote Procedure Call}
\newacronym{rss}{RSS}{Receive-Side Scaling}
\newacronym{rtc}{RTC}{Real Time Clock}
\newacronym{sata}{SATA}{Serial ATA}
\newacronym{sgx}{SGX}{Software Guard Extensions}
\newacronym{smbus}{SMBus}{System Management Bus}
\newacronym{smm}{SMM}{System Management Mode}
\newacronym{smmu}{SMMU}{System Memory Management Unit}
\newacronym{smc}{SMC}{Secure Monitor Call}
\newacronym{soc}{SoC}{System-on-Chip}
\newacronym{som}{SoM}{System-on-Module}
\newacronym{spl}{SPL}{System Protocol Layer}
\newacronym{tap}{TAP}{Test Access Port}
\newacronym{tdp}{TDP}{Thermal Design Power}
\newacronym{tfa}{TF-A}{Trusted Firmware-A}
\newacronym{tlb}{TLB}{Translation Lookaside Buffer}
\newacronym{tpu}{TPU}{Tensor Processing Unit}
\newacronym{ttbr}{TTBR}{Translation Table Base Register}
\newacronym{uart}{UART}{universal asynchronous receiver-transmitter}
\newacronym{uefi}{UEFI}{Unified Extensible Firmware Interface}
\newacronym{upi}{UPI}{Universal Path Interconnect}
\newacronym{vfpga}{vFPGA}{Virtual FPGA}
\newacronym{vpu}{VPU}{Video Processing Unit}
\newacronym{xmpu}{XMPU}{Xilinx Memory Protection Unit}
\newacronym{xppu}{XPPU}{Xilinx Peripheral Protection Unit}

\usepackage{enumitem}
\usepackage{cleveref}

\newcommand\tryagain{\textsc{TryAgain}\xspace}
\newcommand\retire{\textsc{Retire}\xspace}
\newcommand\sysname{\textsc{Lauberhorn}\xspace}

\newcommand*\circled[1]{\raisebox{.5pt}{\textcircled{\raisebox{-.9pt} {#1}}}}

\copyrightyear{2025}
\acmYear{2025}
\setcopyright{cc}
\setcctype{by}
\acmConference[HOTOS '25]{Workshop on Hot Topics in Operating Systems}{May 14--16, 2025}{Banff, AB, Canada}
\acmBooktitle{Workshop on Hot Topics in Operating Systems (HOTOS '25), May 14--16, 2025, Banff, AB, Canada}
\acmDOI{10.1145/3713082.3730388}
\acmISBN{979-8-4007-1475-7/2025/05}

\begin{document}

\title{The NIC should be part of the OS.}

\author{Pengcheng Xu}
\affiliation{
  \institution{ETH Zurich}
  \country{Switzerland}
}
\email{pengcheng.xu@inf.ethz.ch}
\author{Timothy Roscoe}
\affiliation{
  \institution{ETH Zurich}
  \country{Switzerland}
}
\email{troscoe@inf.ethz.ch}

\begin{abstract}

The \gls{nic} is a critical component of a cloud server occupying a
unique position.  Not only is network performance vital to 
efficient operation of the machine, but unlike
compute accelerators like GPUs, the network
subsystem must react to unpredictable events like the arrival of a
network packet and communicate with the appropriate application end
point with minimal latency.

Current approaches to server stacks navigate a trade-off between
flexibility, efficiency, and performance: the fastest kernel-bypass
approaches dedicate cores to applications, busy-wait on receive queues,
etc. while more flexible approaches appropriate to more dynamic
workload mixes incur much greater software overhead on the data path.

However, we reject this trade-off, which we ascribe to an arbitrary (and
sub-optimal) split in system state between the OS and the \gls{nic}.
Instead, by exploiting the properties of cache-coherent interconnects
and integrating the \gls{nic} closely with the OS kernel, we can
achieve something surprising: performance for RPC workloads
\emph{better than the fastest kernel-bypass approaches} without
sacrificing the robustness and dynamic adaptation of kernel-based
network subsystems.

\end{abstract}



\maketitle

\section{Introduction}
\label{sec:intro}

The \gls{nic} is central to the operation of a
data center server, responsible for all performance-critical
communication with the machine.  It differs from other devices in
a critical aspect: unpredictable events (packets arriving)
happen to it during normal operation.  This is in contrast to
components like GPUs (to which the OS submits application tasks with
relatively predictable behavior) or local storage devices (where the
OS issues a request, assuming a response will arrive).

Networking in modern servers is notable for a clear partition of
networking state between the OS, application, and \gls{nic}: broadly,
the \gls{nic} holds state for (de)multiplexing flows, while the OS
holds state related to scheduling of tasks across cores.  This has
evolved over many years and modern \glspl{nic} (including new
architectures proposed in research) encode a number of implicit and
explicit assumptions about trust, OS design, and applications which we
survey in \autoref{sec:trad}.

Surprisingly, these assumptions are preserved in high-performance
kernel-bypass architectures which attempt to bring the \gls{nic}
closer to the application, or CPU designs which integrate the
\gls{nic} with the processor cores.  These techniques deliver
gains in performance, but at the cost of flexibility: bypass generally
relies on a relatively fixed assignment of processes to cores and
queues together with busy-waiting to achieve higher speeds.  This
works well for fairly static workloads, but has limited applicability
for more dynamic application mixes.

Our focus in this paper is on the network receive path, although it is
also closely connected to the transmit path. We also focus on
\glspl{rpc}, whether data center microservices or
serverless function invocations.  While some are large, the great
majority of \gls{rpc} requests and responses are
small~\cite{seemakhupt_cloud-scale_2023}.
Our goal is to exploit this insight to reduce the CPU cycle overhead
of a small \gls{rpc} call \emph{to essentially zero} for many
workloads, within an architecture that nevertheless supports
dynamic workloads with better performance than modern kernel-based
stacks.

While the end-to-end latency of \glspl{rpc} is dominated by
propagation time, \emph{end-system latency} reflects CPU cycles
consumed by the invocation and is therefore a good measurable proxy
for the efficiency of the software stack (unmarshaling,
demultiplexing, function dispatch, etc.).

\begin{figure*}[t!]
    \includegraphics[width=.85\linewidth]{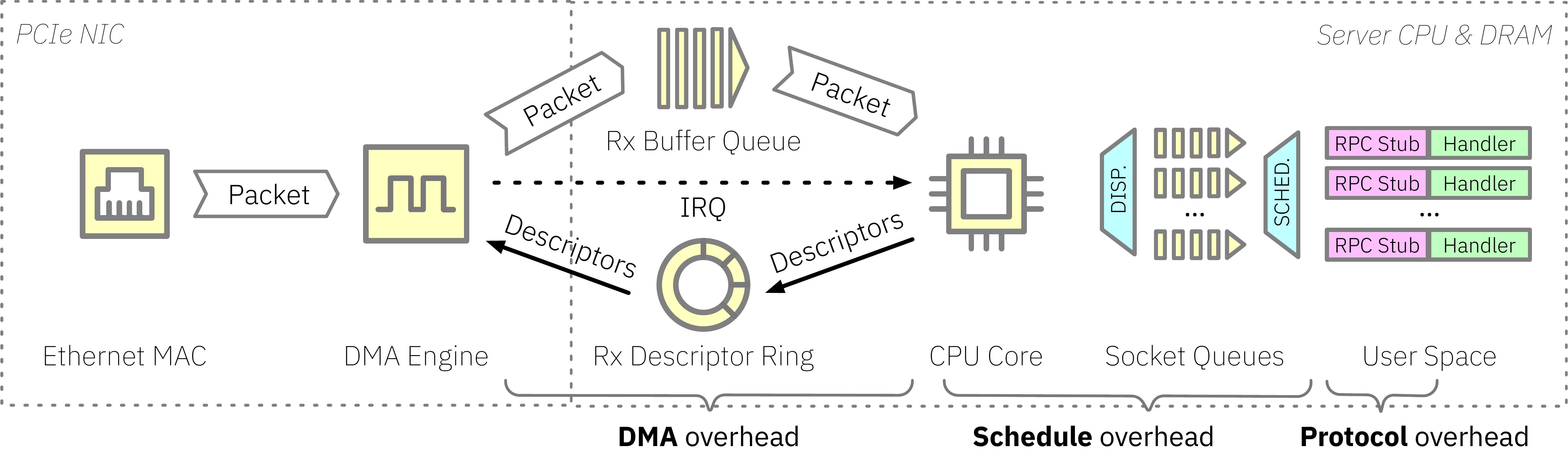}
    \caption{Architecture of a traditional PCIe DMA \gls{nic}'s receive path.}
    \label{fig:ps:pcie_nic}
\end{figure*}

We suggest that kernel-bypass optimization is at its limit,
and argue for a radically different, more OS-centric approach
where the \gls{nic} is a full, trusted component \emph{of the OS
itself}.

Our approach combines several new ideas.  First, we exploit
\emph{cache-coherent peripheral interconnects} and well-established
techniques for protocol offload to transfer data between user-mode CPU
registers and the \gls{nic} with no memory access overhead and no
energy wasted in spinning.  Second, we use the same fine-grained
mechanism to have the kernel \emph{keep the \gls{nic} updated with the
current OS scheduling state}, allowing the \gls{nic} to always steer
packets to the correct user-space process.  Finally, the \gls{nic}
gathers load information and \emph{requests the OS to reschedule
processes} in response to new packets arriving over the network, via
the same lightweight mechanism.

We explore what this means for hardware and software, and the
prototype \gls{nic}, \sysname, that we are building on the Enzian
research platform to demonstrate the idea.

\section{The traditional NIC paradigm}
\label{sec:trad}

Most server network stacks, and also most \gls{nic} hardware designs, are based
around the model in \Cref{fig:ps:pcie_nic}: incoming packets are
demultiplexed and transferred using \gls{dma} into one of a set of
descriptor-based queues, with interrupts used for synchronization when
the OS has stopped polling the queue.  \gls{dma} occurs using
addresses that are translated (and protected) via an \gls{iommu} or
\gls{smmu}, and interrupts can be steered to cores using the
demultiplexing information, or some other heuristic.

This model has evolved over decades from the very simplistic model of
Ethernet interface used in the Xerox Alto, to handle 400Gb/s network
links connected to machines with 100s of cores.  We discuss recent
alternatives below, but observe that most kernel bypass approaches
still look like \Cref{fig:ps:pcie_nic}, but move some parts from the
OS kernel to application user space.  In more detail, a minimal set of
things have to happen to turn a network packet into a function
invocation on the host is:
\begin{enumerate}
  \item \label{st:packetread} Read the packet contents.
  \item \label{st:protocol} Perform protocol processing (checksums, etc.).
  \item \label{st:demux} Demultiplex the packet to an
    in-memory queue.
  \item \label{st:interrupt} Interrupt some CPU core to
    notify the OS, hypervisor, or guest OS.
  \item \label{st:osprot} Perform some general protocol processing.
  \item \label{st:idproc} Identify the OS process (or thread, or task)
    that should handle the message.
  \item \label{st:idcore} Find a (perhaps different) core to execute this process.
  \item \label{st:schedule} Schedule the process on the core.
  \item \label{st:context} Context switch to the process if needed.
  \item \label{st:unmarshal} Unmarshal/deserialize arguments and
    function name.
  \item \label{st:fnaddr} Find the address of the start of
    the function.
  \item \label{st:jump} Jump to this instruction.
\end{enumerate}

A typical \gls{nic} performs steps \ref{st:packetread} to
\ref{st:interrupt}, and then hands things over to the OS or
application.

\textbf{Kernel-bypass} approaches like Arrakis~\cite{peter_osdi_2014},
IX~\cite{ix}, and Demikernel~\cite{zhang_demikernel_2021} variously trade off
latency and throughput against flexibility and energy efficiency by
replacing step \ref{st:interrupt} with spinning or polling, and simplifying
steps \ref{st:osprot} through \ref{st:context} by
binding application processes to in-memory queues in advance.  The
data plane is moved to application space, while the control plane can
be left in the OS kernel or moved to dedicated cores
(Shinjuku~\cite{kaffes_shinjuku_2019},
Shenango~\cite{ousterhout_shenango_2019},
Caladan~\cite{fried_caladan_2020}) or userspace processes
(ghOSt~\cite{humphries_ghost_2021}).  Snap~\cite{google-snap},
meanwhile, dedicates a subset of the CPU cores to provide
applications a uniform, yet highly configurable, abstraction of a
\gls{nic} that allows rapid deployment of new network stack features.

All these approaches, however, retain the same division of labor
between software and the \gls{nic}; indeed, all of them resemble
\Cref{fig:ps:pcie_nic}. The principal differences concern
the kernel/user space boundary (and where the different receive path
stages execute) and the design of the control plane (which is
implemented in software on the CPU as a separate component).  To a
large extent, kernel bypass turns what was OS functionality into
application-level functionality, \textit{integrating the
  \gls{nic} more with the user application}.

Other work from architecture has explored closely \textbf{integrating
  the \gls{nic} with the CPU}.  nanoPU~\cite{ibanez_nanopu_2021}
delivers packets processed by P4 directly into the register file of a
RISC-V core, while CC-NIC~\cite{schuh_cc-nic_2024} uses a NUMA server
to explore by emulation the implications of cache-coherent peripheral
interconnects for \glspl{nic}.  This, again, preserves the same
hardware/software boundary, while heavily optimizing hardware steps
\ref{st:demux} and \ref{st:interrupt}.

As with bypass, this works well when the workload is
relatively static, can be bound to dedicated cores, and is
rarely idle.  However, when the workload is dynamic with many
more end-points than spare cores,
the up-front cost of mapping the NIC's
demultiplexing to queues onto the scheduling of applications on cores
quickly becomes cumbersome.  Even newer \gls{dpu}~\cite{aws:nitro:security} and
\gls{ipu}~\cite{humphries_tide_2024} systems share these
characteristics.

Moreover, tightly coupling
the \gls{nic} and CPU may not be desirable:  Networking parts do not
develop in lock-step with CPUs and different workloads have
very different compute-to-network I/O ratios, so there is valuable
flexibility gained by keeping the NIC as a separate component.

\section{Why this split?}\label{sec:split}

The historical stability of the hardware/software boundary in
\glspl{nic} is arguably due to the state required to perform each step.
For example, steps \ref{st:unmarshal}-\ref{st:jump} require
application-specific state: argument formats, interface signatures,
and code layout.  In contrast, steps \ref{st:osprot}-\ref{st:context}
cannot be performed without reference to central OS state.

A key factor is that \textbf{the OS doesn't trust the \gls{nic}}.
  Kernel developers have been keen to limit the coupling between
  OS and \gls{nic}~\cite{mogul_tcp_offload} due to the perception that
the \gls{nic} never does quite what the OS designer wants.
Ironically, the result continues to be an increase of the complexity
of device \emph{drivers} as hardware vendors adopt ad-hoc solutions to
exposing functionality to users.  This in turn means that the
functionality that the vendors add to a NIC is limited to \emph{that
which can easily be exposed to users}.

Moreover the introduction of \glspl{iommu} and \glspl{smmu} has led to
a philosophy that, as far as possible the \gls{nic} should not be
trusted as a device.  This is an anomaly, given that devices like
disks, CPU cores, GPUs, and DRAM are, for the most part, trusted by at
least part of the OS.  One reason for this is confusion about
different roles of the \gls{smmu}: on the one hand, providing a
convenient memory translation function on the data path to facilitate
device pass-through to virtual machines, and on the other, to firewall
off a kernel running on a set of application cores from the rest of
the machine.

The philosophy is compounded by protocols like RDMA which regard the
\gls{nic} as a relatively dumb device with little connection to the
host OS that can nevertheless perform memory accesses on behalf of a
remote peer, using an authorization framework that is naive at best in
multi-tenant scenarios. 
A more OS-centric perspective on RDMA-like
functionality views the \gls{nic} as providing \textit{to the OS}
additional, specialized cores close to the network interface which can
execute a limited number of \gls{rpc} operations.

A related factor is that \textbf{architecture researchers like to
  ignore the OS}~\cite{mogul_hotos_2011}.  User applications are a
different matter, and so there are many proposals for accelerating
subsets of steps \ref{st:unmarshal}-\ref{st:jump} for memory latency
benefits.  Cereal~\cite{jang_specialized_2020} proposed an accelerator
targeting a custom message format; the accelerator sits directly on
the system interconnect \emph{inside} the CPU package.  Optimus
Prime~\cite{pourhabibi_optimus_2020} proposed a format-agnostic
transformation architecture and focused on implementing an accelerator
sitting on the system interconnect inside the CPU package as well.
Cerebros~\cite{pourhabibi_cerebros_2021} builds upon Optimus Prime
towards a fully-offloading RPC framework.
ProtoAcc~\cite{karandikar_hardware_2021} targets Protocol Buffers
instead with an accelerator attached to the custom
RoCC~\cite{asanovic_rocket_2016} interface directly on the RISC-V core
pipeline.
Like kernel bypass, these primarily target static assignments of
applications to cores and accelerate single application
performance in part by imposing strong assumptions to minimize steps
\ref{st:osprot}-\ref{st:context}.

Underlying this apparent trade-off between performance (static
assignment of cores) and flexibility (more OS involvement) is the
\textbf{misconception that fine-grained interaction between OS and
  \gls{nic} is slow:}  the \gls{nic} is not just untrustworthy but also
hard to talk to.   In cases such as \gls{rss} the goal is to
provide offload (e.g. load balancing) without involving the OS
\textit{at all}.
This assumption may hold for DMA descriptor rings, but much less so for
loads/stores to device registers over modern \gls{pcie}, and even less
the case when accessing the device over new, cache-coherent interconnects
like CXL.mem 3.0.

\section{Breaking the impasse}\label{sec:insights}

\begin{figure}[t]
    \includegraphics[width=0.9\linewidth]{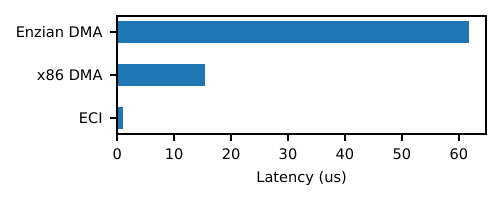}
    \caption{64-byte message round-trip latencies.} \label{fig:lat}
\end{figure}

\begin{figure*}[t!]
    \includegraphics[width=.85\linewidth]{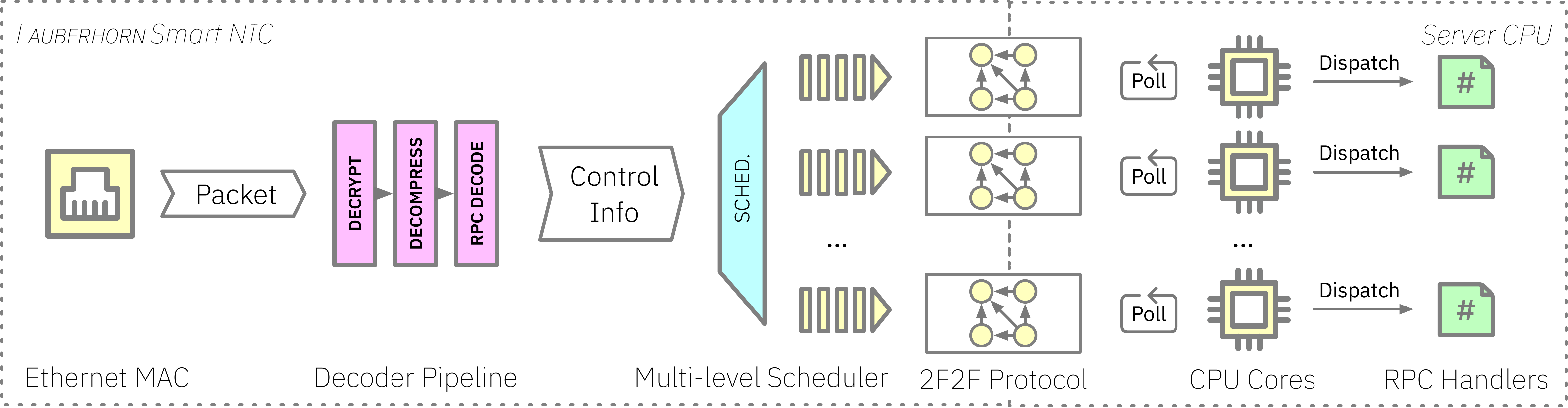}
    \caption{Overview of the \sysname receive path.}
    \label{fig:po:flow}
\end{figure*}

We are pursuing a different approach in order to deliver performance
\emph{better than current kernel bypass} for relatively stable RPC and
serverless workloads, while providing all the flexibility of the
traditional approach with better efficiency for highly dynamic
workloads.  We exploit three new insights.

Firstly, new \textbf{cache-coherent peripheral interconnects} between
devices and cores can radically change communication between CPU and
\gls{nic}.  Examples are CXL.mem 3.0~\cite{cxl_consortium_compute_2022}, CCIX~\cite{ccix},
and the \gls{eci}~\cite{cock-asplos-2022}.  Crucially, this allows
lightweight signaling to the device: a \gls{nic} can interpret cache
operations on certain addresses as specific signals or requests, and
return information back to the CPU in response or trigger other
actions such as interrupts.  \autoref{fig:lat} shows the dramatically
better interaction latency possible using even the 
(comparatively slow) \gls{eci} vs. \gls{dma} over \gls{pcie} on the
same machine, and on a modern PC server; we anticipate comparable
gains with CXL 3.0.

For the data plane, such protocols allow packets to be transferred
directly as cache lines to the destination
core's L1 cache and registers~\cite{ruzhanskaia_rethinking_2024},
providing dramatically lower latency than can be achieved using DMA
with descriptors.

For the control plane, communication with between CPU cores (running
either application code or the OS kernel) is lightweight and
efficient, and easily protected using conventional MMU mechanisms.  It
also conveys rich information, for example, a \gls{nic} can infer
whether a core is polling in user mode or in kernel mode based on
which address is requested from its home address space.

Secondly, \textbf{it's time to trust the \gls{nic}}.  The \gls{nic} is
a critical part of the OS function of the machine.  Unlike, e.g., the
GPU, it is enabling infrastructure for the whole system.  Viewing it
as a potential part of the OS rather than an untrusted peripheral is
the only way to fully exploit its hardware resources and unique
position in the data path.

In particular, since the \gls{nic} is responsible for demultiplexing
an incoming packet to an application end-point, it should have access
to all the relevant OS state: which processes are currently in the run
queues on which cores, which are currently executing, and which are
waiting.  Some of this can be inferred from the cache traffic the
\gls{nic} observes as in the example above, while any other state can be
explicitly pushed to the \gls{nic} via the interconnect with
negligible overhead.

Finally, adopting the previous positions allows us to \textbf{fully
  implement RPC dispatch} on the \gls{nic}.  Integrating existing
techniques for accelerating deserialization with
rich knowledge of the OS state enables RPC dispatch with
essentially zero software overhead: in the common case, it is
possible to execute \emph{every} step in \Cref{sec:trad} on the
\gls{nic}, and have a stalled load on a processor core return a
carefully prepared cache line with only the information needed to
dispatch an RPC: just the arguments and virtual address of the
first instruction of the target function to jump to.

Sharing the OS state means that this efficiency is preserved
when executing dynamic workloads where statically
associating DMA queues, cores, threads, and sockets is not practical:
the \gls{nic} already has information about whether, and where, a target process
is running and can notify either it or the OS accordingly.  Moreover,
the OS has up-to-date information from \sysname about which core are
polling and where, so as to guide scheduling decisions.

\section{Implementation Sketch}
\label{sec:proto-offload}

We are building a prototype, \sysname, to demonstrate the feasibility
of these ideas on the context of microservices.  \sysname exploits the
large FPGA, 100Gb/s interfaces, and cache-coherent interconnect on the
Enzian research computer~\cite{cock-asplos-2022}.  We sketch two key
components of the design: the receive fast path, and the sharing of
scheduling state between the OS and \gls{nic}; we return to
additional, non-functional issues to be addressed in this design in
\autoref{sec:next-steps}.

\subsection{Receive fast path}

\autoref{fig:po:flow} shows an overview of a minimal receive path, in
the case where the receiving process is already executing on at least
one core in the system and at least one execution thread of that
process is ready to process a request.

\sysname demultiplexes and unmarshals an incoming \gls{rpc} request
packet using information provided in advance by the OS kernel (and,
indirectly, the application service) to give (1) a process and
communication end-point, (2) a \textit{code pointer} and \textit{data
  pointer} inside that process corresponding to the request, and the
\textit{call arguments}, corresponding to steps
\ref{st:packetread}-\ref{st:demux}, \ref{st:osprot}-\ref{st:idproc},
\ref{st:unmarshal}, and \ref{st:fnaddr} in \autoref{sec:trad}.
This can be achieved using a variation of existing \gls{nic}
techniques like protocol offload and \gls{rpc} deserialization
acceleration (e.g.~\cite{pourhabibi_optimus_2020}).

In the FPGA implementation, an Ethernet frame streams in from the
Ethernet MAC IP block and passes through various streaming-mode header
decoders to demultiplex the packet and remove the Ethernet, IP, and
UDP headers, storing them in SRAM for any upper protocol layers that
require buffering.

\begin{figure}[t]
    \includegraphics[width=0.9\linewidth]{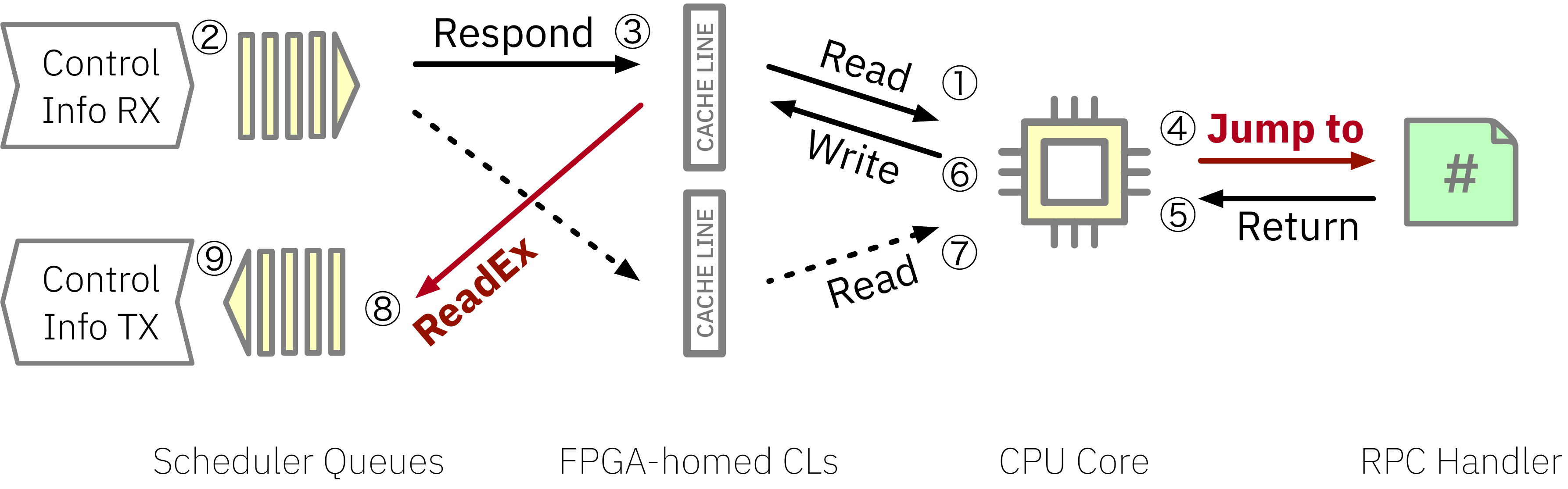}
    \caption{The \sysname protocol between NIC and CPU}
    \label{fig:po:2f2f}
\end{figure}

\sysname then delivers a minimal data structure to the destination
communication end-point which consists of the target code and data
pointers, together with the \gls{rpc} arguments.  It does this using
an extension of the protocol described by Ruzhanskaia \textit{et
al.}~\cite{ruzhanskaia_rethinking_2024} (\Cref{fig:po:2f2f}).  Each
end-point comprises a set of cache lines homed on the \gls{nic}: two
\textsc{control} lines plus multiple \textsc{auxiliary} lines to
handle payloads larger than a single cache line (128\,B on Enzian).
The transmit path uses a similar, disjoint set of cache lines.

\begin{figure*}[t]
    \includegraphics[width=0.9\linewidth]{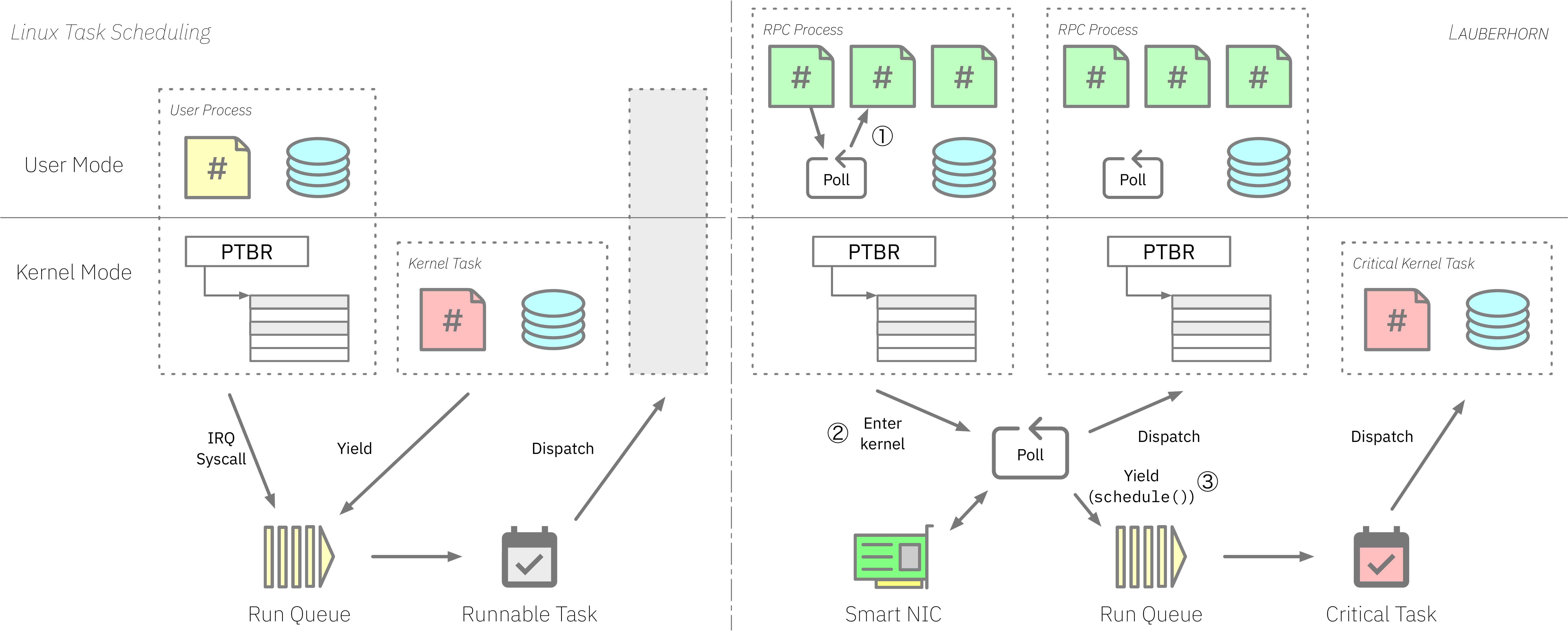}
    \caption{Comparison between normal task scheduling and NIC-driven
    scheduling of RPC isolation domains.} \label{fig:sched:comparison}
\end{figure*}

To receive a request, a process issues a load to one
\textsc{control}
cache line to receive a request.   The \gls{nic} with respond to this
load with the data listed above when an appropriate packet arrives and
has been decoded; until then the core is stalled (rather than
spinning).

The CPU now has all the information it needs to start executing the
first instruction of the user procedure in registers and L1 cache, and
is already in the correct address space.  It executes the \gls{rpc}
handler and writes the \gls{rpc} result into the same \textsc{control} cache line, and loads the second \textsc{control} line for
the next packet.  \sysname sees the load for the second line and thus
knows that the CPU has finished serving the first request.  Before
responding to the read on the second cache line, the \gls{nic} issues
a \emph{fetch exclusive} over coherence protocol to request the first
line (containing the \gls{rpc} response) from the CPU's cache  and
sends it out over the network.  Finally, when the next packet arrives
and is decoded, \sysname responds to the CPU's read on the
second \textsc{control} cache line.

Of course, \sysname cannot block a cache fill from a core
\emph{indefinitely} without a timeout in the coherence protocol
causing an unrecoverable ``bus error'' which leaves the
system in an inconsistent state.  We avoid this by
returning \tryagain dummy messages after 15ms, reducing the polling
overhead (both bus traffic and CPU spinning) to almost zero and
improving energy efficiency.

Moreover, this provides a mechanism for cleanly descheduling the
process: while the OS can still preempt a running process at any time,
a core blocked on such a communication load provides a useful
synchronization point.  \sysname can notify the OS that the
process has blocked, the OS (or the \gls{nic}) can send an \gls{ipi} to
the process' core, and then \sysname can send the process a \tryagain
message, unblocking it and causing to immediately enter the kernel.

\subsection{Demultiplexing and scheduling}
\label{sec:sched-integration}

A key novelty of \sysname is how it uses
precise kernel scheduling state to dispatch requests.
In the fast case, a request arrives directly at the correct
process without kernel intervention, since \sysname is aware which
core is running the process and waiting for a cache line
holding the request.  Efficiently keeping this state up to date across
context switches is practical in part due to the extremely
low latency of communication between the CPU and \gls{nic} enabled by
the interconnect. 

When no core is running the destination process for a received packet,
\sysname quickly delivers the request to the kernel, allowing it 
schedule the target process and deliver the unmarshalled request in
software.  \Cref{fig:sched:comparison} compares this approach to the
traditional Linux dispatch loop.

A CPU core running a regular kernel thread uses the
protocol in \Cref{sec:proto-offload} to monitor a pair of
\textsc{control} cache lines for incoming requests; \sysname can 
dispatch a request for \emph{any} process to this CPU core and
end-point, whereupon the CPU switches to the corresponding process to
handle the request. 
As it is a conventional kernel thread, it periodically calls
\texttt{schedule()} \circled{3} and can handle regular critical
kernel operations like Read-Copy-Update.

Thereafter \circled{1} the core remains in the same process
and runs a user-mode loop on a \emph{different} pair of \textsc{control}
cache lines which \sysname has dedicated to that process.  At this
point, dispatching requests to this service involves almost no
software overhead: the load executed by the core immediately returns
the address to jump to.

This works well under the assumption that the number of ``hot''
services is less than the number of available cores -- 48 on Enzian,
and often in the hundreds for modern server-class processors.

The user-mode loop can give up the CPU in a variety of ways
\circled{2}.  The process can voluntarily yield the CPU by executing a
system call whenever it is executing; in the case that the core is
blocked on a load of a \textsc{control} cache line, this will occur
when it receives an \gls{rpc} payload or a \tryagain message from
\sysname.  Alternatively, the kernel and \sysname can cooperate to
fully \emph{preempt} the user process by sending an interrupt to the
core, and then resuming it (allowing it to receive the interrupt) with
a subsequent \tryagain as described above. Note that this can be
initiated by the kernel scheduler, or by \sysname based on statistics
it gathers about the instantaneous load on each server process.  This
approach therefore also supports dynamic scaling of the cores used for
\gls{rpc} based on load.

Many data center deployments use non-preemptive kernels for
throughput.  \sysname provides dynamic load information to the kernel
(using, again, the kernel-mode control channels) to reallocate cores
between \gls{rpc} services and non-\gls{rpc} processes.  Much as we
already preempt user-space threads blocked waiting for a cache line,
any non-preemptable kernel thread waiting on \sysname can be
reallocated by sending it a \retire message from the \gls{nic}.

\section{Open questions and concerns}
\label{sec:next-steps}

The fine-grained concurrent interaction in \sysname between
application threads, OS kernel processes, the cache coherence
protocol, and the \gls{nic} itself is subtle, and correct operation of
the system requires us to ensure that all races are benign.
Fortunately, we have found that the problem is highly amenable to 
specification using TLA+, and can be model-checked for correctness
relatively easily. 

\sysname as described so far will support full-featured \gls{rpc}
interaction with high efficiency, but is lacking some non-functional
features that become important in real data center settings. While
encryption can be handled with fairly standard techniques, support for
tracing, debugging, and statistics presents interesting properties for
further close integration with the OS.

For large messages, the direct, low-latency approach becomes less
efficient and it is best to revert back to DMA-based transfers since
throughput comes to dominate over latency.  The trade-off will depend
on the platform, empirically for Enzian this happens at about 4KiB.

Nested \glspl{rpc} will benefit from the ability to rapidly create a
dedicated end-point for an \gls{rpc} reply.  Fine-grained interaction
with the \gls{nic} should make creating this continuation a cheap
operation with significant performance benefits. 

The design of standard OS-\gls{nic} and application-\gls{nic}
interface is an open question, one which we hope to answer through
building \sysname as a prototype thus evolving the interfaces we
provide.  

\section{Acknowledgements}
\label{sec:acks}

We thank the anonymous reviewers for their constructive comments on
the paper, and the rest of the Enzian team for their support and
ideas.  This work was partly funded by a gift from the Google Systems
Research Group, for which we are grateful.

\bibliographystyle{acm}
\bibliography{ref}

\end{document}